\documentclass{scrartcl}
\usepackage[utf8]{inputenc}

\title{Measuring Earth's Magnetic Field Using a Smartphone Magnetometer}
\author{Simen Hellesund \\ University of Oslo}
\date{}

\usepackage{graphicx}
\usepackage{multicol}
\usepackage[T1]{fontenc}
\usepackage{float}
\usepackage{amsmath}
\usepackage{upgreek} %roman greek characters
\usepackage{hyperref}
\usepackage{url}
\usepackage{cite} %dashes when citing several papers

\usepackage{dcolumn}
\usepackage{booktabs}

\begin{document}\sloppy

\twocolumn[{%
\maketitle
\section*{Abstract}
I use the magnetometers of a smartphone to measure the magnitude of the earth's magnetic field. A method for eliminating bias due to internal magnetic fields in the phone is proposed. \vspace{.3em}

\textit{Keywords}: Earth's magnetic field, magnetometer, Hall-effect.

\vspace{5mm}
}]

\section*{Introduction}
In recent years, modern smartphones have become packed with sensors: Accelerometers, gyroscopes, microphones, cameras, proximity sensors, and more. Most people are walking around with a potential laboratory in their pocket, the like of which would have been the envy of any experimenter in times gone by. In recent years, numerous papers have been published describing smartphone-based experiments for educational purposes, utilising the various sensors of the smartphone\cite{cit:Hockberg,cit:Pierratos,cit:Kuhn,cit:Kuhn2,cit:Kuhn3,cit:Sans,cit:Castro}. 

Magnetometers, sensors for measuring magnetic fields, have become ubiquitous in smartphones. These magnetometers are what enables the phone's compass to function. Several papers have also been published describing how such sensors could be used in education\cite{cit:Arribas,cit:Arabasi,cit:Setiawan,cit:Septianto}.

The magnetometers used in smartphones utilise the so-called \textit{Hall-effect} to operate. This effect was first demonstrated by Edwin Hall in 1879. When a current is flowing through a thin foil of semiconductor material, a voltage difference will arise perpendicular to the direction of the current in the presence of a magnetic field perpendicular to the semiconductor foil\cite{cit:Hall}. This voltage is linearly dependent on the strength of the magnetic field. Today, the material needed to make a Hall-sensor sensitive enough to measure the strength of the earth's magnetic field is small enough to fit inside a mobile phone. Most phones will, in fact, contain three magnetometers, one for each spatial direction. 

There are a plethora of apps available for accessing the instantaneous readings of a phone's magnetometer. For this experiment, the app named \textit{Physics Toolbox Magnetometer}\cite{cit:Magnetometer}, available in the Google Play store, is used. The simplest way of measuring the field strength of the earth's magnetic field would be to simply take the vector sum of the field strength contributions in each spatial direction. This is done automatically in the Physics Toolbox Magnetometer app. However, when testing the app with the smartphone used in this experiment, I observe that the physical orientation of the phone has a noticeable effect on the field strength readings. This indicates the presence of systematic offsets in the magnetometer sensors, or constant magnetisation of the phone. I suggest a method to eliminate these effects, thus achieving a more accurate measurement. 

\section*{Experimental Setup}
\begin{figure}[ht]
    \centering
    \includegraphics[width=.4\textwidth]{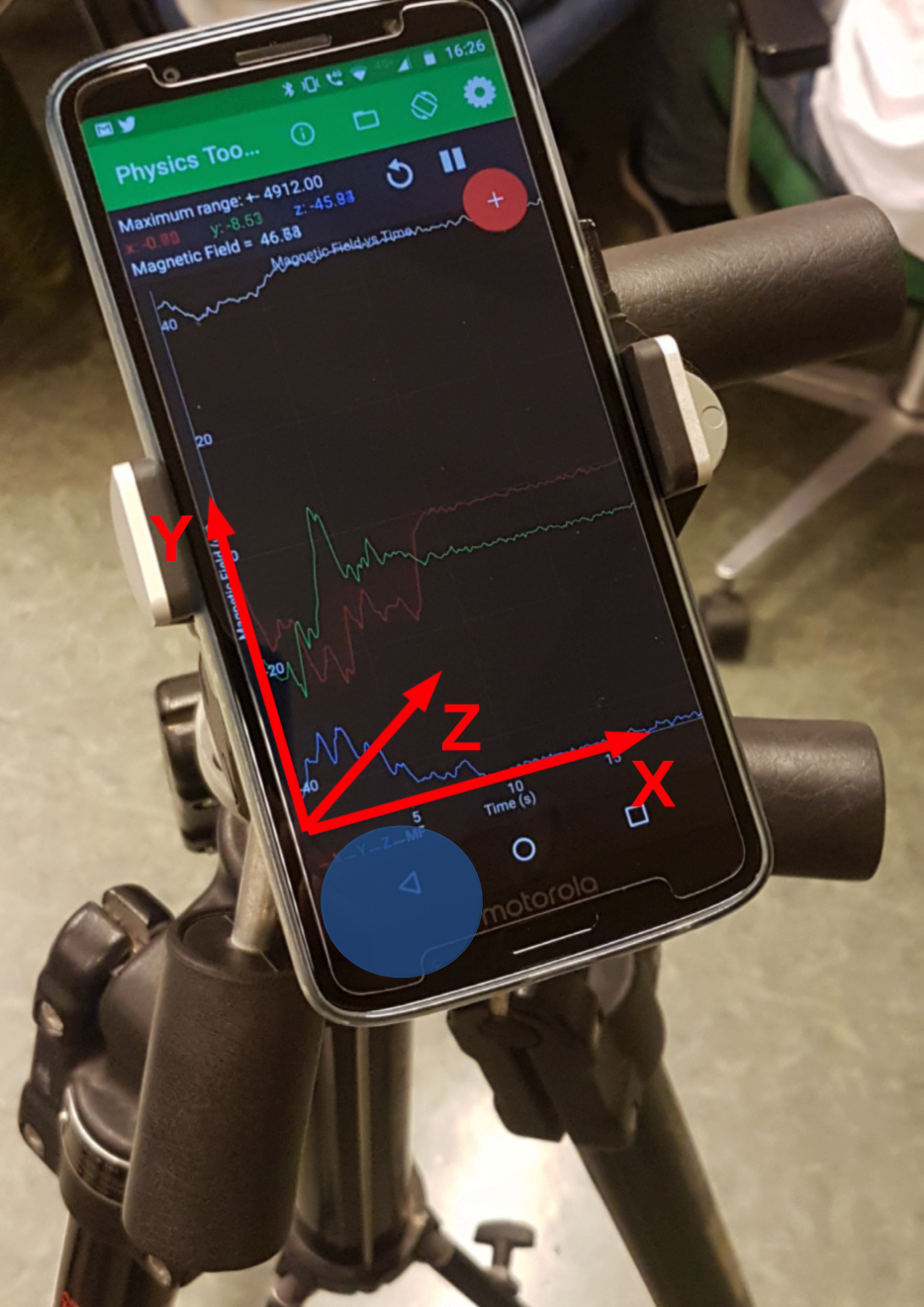}
    \caption{Experimental setup. The tripod seen in the picture was not used in the experiment.}
    \label{fig:Setup}
\end{figure}

The only equipment needed to perform the experiment is a smartphone with a magnetometer app installed. The phone used here is a Motorola Moto g6 Plus, shown in Figure \ref{fig:Setup}. The approximate location of the magnetometer sensors in the phone is determined by moving a test magnet across the surface of the phone. The magnetometer reading spikes at the closest approach of the test magnet to the magnetometer. The approximate location of the sensors is marked in blue in the Figure. This location is not critical to the experiment.

The coordinate system of the magnetometer app is as follows: The positive x-direction is defined as pointing directly left to right in the plane of the phone screen. The y-direction also lies in this plane and is pointing directly upwards along the screen. The z-direction is perpendicular to the x-y-plane and is pointing directly out of the screen. This coordinate system is drawn in Figure \ref{fig:Setup}.

\section*{Procedure}
As mentioned in the previous section, the magnetic field strength reading of the magnetometer app is found to be quite strongly dependent on the orientation of the smartphone. This means that the magnetic field measured by the magnetometer will be the sum of the earth's magnetic field $B_{\mathrm{earth}}$ and the constant magnetic offset $B_{\mathrm{offs}}$ in the phone:  
\begin{equation}
    B_{\mathrm{meas}} = B_{\mathrm{earth}} + B_{\mathrm{offs}} \,.
    \label{eq:B1}
\end{equation}
If the phone is flipped such that the magnetic field is measured in the opposite direction, the offset term will be the same, but the sign on the contribution from the earth's magnetic field will change:
\begin{equation}
    B_{\mathrm{meas}}^{\pi} = -B_{\mathrm{earth}} + B_{\mathrm{offs}} \,.
    \label{eq:B2}
\end{equation}
The constant offset $B_{\mathrm{offs}}$ can be eliminated by adding Equations \ref{eq:B1} and \ref{eq:B2}. The magnetic field of the earth is then taken as the mean of the two measurements:
\begin{equation}
    B_{\mathrm{earth}} = \frac{B_{\mathrm{meas}} + B_{\mathrm{meas}}^{\pi}}{2} \,.
\end{equation}

The smartphone contains three magnetometers, one for each spatial direction. To minimise the effect of potential miscalibration of any one of the sensors, the measurement is performed individually for each sensor. This is done by orienting the smartphone such that only the magnetometer in question has a reading.

I was not able to find any information on the resolution of the analog to digital converter of the magnetometers. This uncertainty is assumed to be lower than the readout uncertainty of the measurement; as I am not able to hold the phone completely still during the measurement, the field strength reading fluctuates somewhat. The field strength is thus determined to the nearest $\upmu$T. 

The experiment was first performed indoors. This yielded inconsistent results, possibly due to a permanent magnetisation of the building in which the experiment was done. Iron rebars in the concrete of the building, for example, may have obtained a permanent magnetisation. The experiment was therefore moved outside.

The experiment was also first performed by attaching the phone to a camera tripod, using a clamp. The tripod allows for accurate orientation of the phone in all directions. The tripod used can be seen in Figure \ref{fig:Setup}. However, the metal legs of the tripod were found to be magnetised. This magnetisation interfered with the measurements. The tripod was therefore abandoned in favour of simply orientating the phone by hand.

\section*{Results}
The results of the measurements can be found in Table \ref{tab:results}. Two measurements are performed for each magnetometer according to the procedure described in the previous section. Taking the average of each such pair of measurement removes the constant offset due to internal magnetic fields in the phone. This gives one value for the magnetic field strength of the earth for each magnetometer. These values are stored in Table \ref{tab:results2}. Taking the average of these three measurements gives the final measurement of the earth's magnetic field: B$_{\mathrm{earth}} = 50.5 \pm 0.4 $ $\upmu$T. The uncertainty on the measurement is found by adding the uncertainties on the individual measurements in quadrature. The field strength obtained is reasonable for Oslo, Norway, where the experiment is performed.

\begin{table}[ht]
\centering
\begin{tabular}{l D{,}{\pm}{-1}}
\toprule
Direction     &  \multicolumn{1}{c}{$B_{\mathrm{meas}}$ [$\upmu$T]} \\ \hline
x     & 47,1      \\
x$^{\pi}$ &-53,1  \\
y     & 51,1      \\
y$^{\pi}$  & -52,1 \\
z     & 51,1      \\
z$^{\pi}$  & -49,1 \\
\bottomrule
\end{tabular}
\caption{Magnetic field strength measurements.}
\label{tab:results}
\end{table}

\begin{table}[ht]
\centering
\begin{tabular}{ll}
\toprule
Measurement (dir.) &  $B$\textsubscript{earth}  [$\upmu$T] \\ \hline
1 (x)     & 50.0$\pm$0.7       \\
2 (y)     & 51.5$\pm$0.7       \\
3 (z)     & 50.0$\pm$0.7       \\
\bottomrule
\end{tabular}
\caption{Magnetic field strength measurement. Averaging out constant offsets.}
\label{tab:results2}
\end{table}

\section*{Conclusion}
In this article, I demonstrate an economic experiment for measuring the earth's magnetic field. The only equipment needed is a smartphone with magnetometer sensors. A method is shown to eliminate the effect of internal magnetic fields in the phone.

Students taking an experimental course on electromagnetism can, for example, use this method along with other methods for measuring the earth's magnetic field, to check their results. The experiment also provides an opportunity for students to practice estimating uncertainties and systematic errors.

Steps should be taken to perform the experiment away from any potential sources of magnetic contamination. 

\bibliographystyle{unsrt}
\bibliography{biblio.bib}

\end{document}